\documentclass[aps,prd,12point,twocolumn,nofootinbib,showpacs,superscriptaddress]{revtex4-1}
\def\theequation{\arabic{section}.\arabic{equation}}
\usepackage{psfrag}
\usepackage{subfigure}
\usepackage{color}
\usepackage{mathrsfs}
\usepackage{graphicx}
\usepackage{amssymb, bm}
\usepackage{amsmath, amsthm}
\usepackage{epstopdf}
\usepackage{hyperref}
\usepackage{enumerate}
\usepackage{longtable}
\usepackage{amsmath}

\newcommand{\be}{\begin{equation}}
\newcommand{\ee}{\end{equation}}

\begin{document}
\def\theequation{\arabic{section}.\arabic{equation}}

\title{Geometry of static $w=-1/5$ perfect fluid spheres in general
relativity}



\author{Behnaz Fazlpour}
\email[b.fazlpour@umz.ac.ir]{}
\affiliation{Department of Physics, Babol Branch, Islamic Azad
University, Babol, Iran}

\author{Ali Banijamali}
\email[a.banijamali@nit.ac.ir]{}
\affiliation{Department of Basic Sciences, Babol Noshirvani
University of Technology, Babol, Iran}

\author{Valerio Faraoni}
\email[vfaraoni@ubishops.ca]{}
\affiliation{Department of Physics \& Astronomy, Bishop's
University, 2600 College Street, Sherbrooke, Qu\'ebec, Canada
J1M~1Z7}


\date{\today}

\begin{abstract}

We discuss the physical features of two recent classes of analytical
solutions of the Einstein equations sourced by an exotic perfect fluid
with equation of state $ P=-\rho/5$. These geometries depend on up to
four parameters and are static and spherically symmetric. They describe
compact spaces with naked central singularities.

\end{abstract}


\maketitle

\section{Introduction}
\label{sec:1} \setcounter{equation}{0}

Recently, two new families of static and spherically symmetric solutions of the
Einstein equations (without cosmological constant) were proposed by Semiz
\cite{Semiz:2020lxj}. The matter source is a perfect fluid with constant
barotropic equation of state $ P =-\rho/5 $, where $\rho$ and $P$ are the fluid
energy density and pressure, respectively \cite{Semiz:2020lxj}. One would like
to understand the physical nature of these solutions and assess whether they
can be useful to model regions of stars, at least as toy models. The equation
of state $P=-\rho/5$ is clearly unphysical, as one would be hard put to find
realistic situations described by this fluid, but dark energy-like stars
(and even phantom energy stars \cite{DeBenedictis:2008qm}) have
been studied in the literature \cite{Chapline:2004jfp, Lobo:2005uf,
Bilic:2005sn, Chan:2008rk, Yazadjiev:2011sm, Rahaman:2011hd, Horvat:2012aq,
Bhar21}, as well as halos of exotic energy \cite{Armendariz-Picon:2005oog}.
Although dark energy has pressure $P<-\rho/3$ and there are all indications
that, if it is responsible for the present acceleration of our
universe, it has equation of state $P \simeq -\rho $
\cite{AmendolaTsujikawabook}, our situation with
$P=-\rho/5$ could still serve as a toy model for hypothetical objects
formed by a
negative pressure fluid. Moreover, from the
mathematical point of view, simple
solutions of the Einstein equations describing perfect fluids are relatively
difficult to find. Although there are over one hundred analytical solutions of
the Einstein equations sourced by perfect (and even imperfect) fluids
that constitute
potential candidates to model relativistic stars, or at least stellar
regions
\cite{Stephani,Delgaty:1998uy}, almost all of them turn out to be unphysical
for one reason or another \cite{Delgaty:1998uy}. Here we examine the new
solutions of \cite{Semiz:2020lxj} to understand their physical features (or
lack thereof).  These geometries are written in Buchdahl coordinates but it is
more instructive from the physical point of view to rewrite them in terms of
Schwarzschild-like coordinates, which we do here.

We follow the notation of Ref.~\cite{Waldbook}: the metric signature
is ${-}{+}{+}{+}$ and we use units in which the speed of light in
vacuo $c$ and Newton's constant $G$ are unity, while $ \kappa \equiv 8\pi
G$ to keep with Ref.~\cite{Semiz:2020lxj}.

Semiz's proposal consists of a four-parameter family of solutions of
the Einstein equations with zero cosmological constant
\be
{\cal R}_{ab}-\frac{1}{2} \, g_{ab} {\cal R} = \kappa  \,T_{ab} \,,
\ee
where ${\cal R}_{ab}$ is the Ricci tensor of the metric $g_{ab}$ and
${\cal R}$ is the Ricci scalar. The matter source is a perfect fluid
with stress-energy tensor
\be
T_{ab}= \left( P+\rho \right) u_a u_b +P g_{ab} \,,
\ee
where  $u^a$ is the fluid 4-velocity and the equation of state is
$P=-\rho/5$ \cite{Semiz:2020lxj}. These geometries are spherically
symmetric and static in the appropriate coordinate range. There are
two new classes of solutions in \cite{Semiz:2020lxj}: the most
general family is parametrized by four constants $\left( C_0,
C_1, C_2, C_3 \right)$ and has line element
\begin{eqnarray}
ds^2 &=& -\frac{3C_1 \left( C_0 +C_1r \right)}{f(r)} \, dt^2
+\frac{f(r)}{3C_1 \left( C_0+C_1r \right)} \, dr^2 \nonumber\\
&&\nonumber\\
&\, & + \frac{f^2(r)}{9C_1^2} \, d\Omega_{(2)}^2 \,,
\label{generalle}
\end{eqnarray}
with $C_1\neq 0$, $ C_0 +C_1r \neq 0$, and where
\be
f(r) = 3\left(
C_1 C_2 +r \right)+C_3 \left( C_0 +C_1r \right)^3 \,,
\ee
while $d\Omega_{(2)}^2 \equiv d\vartheta^2 + \sin^2 \vartheta \,
d\varphi^2$ is the line element on the unit 2-sphere. The energy
density is \cite{Semiz:2020lxj}
\be
\rho(r) = -5P(r)=  \frac{-45 \kappa \,
C_1^3 C_3 \left( C_0 +C_1 r \right)^2}{f^2(r)}   \label{rho}
\ee
and is non-negative provided that
\be
C_1 C_3 \leq 0 \,,
\ee
which we assume in the following, while the limiting situation given by
$C_3=0$ corresponds to vacuum. The solution for $C_1=0$ is not obtained
continuously from Eqs.~(\ref{generalle}) and (\ref{rho}) in the limit
$C_1\to 0$ but requires a separate discussion \cite{Semiz:2020lxj}. This
second family is parametrized by the remaining three constants
\cite{Semiz:2020lxj}: we begin by analyzing this second family (or
``special solution'' in the nomenclature of \cite{Semiz:2020lxj}) in the
following section.

\section{Special solution $C_1=0$}
\label{sec:2} \setcounter{equation}{0}

This 3-parameter $\left( C_0, C_2, C_3 \right)$ family of solutions
is described by the line element \cite{Semiz:2020lxj}
\be
ds^2=-\frac{C_0^2}{g(r)} \, dt^2 + \frac{g(r)}{C_0^2} \, dr^2 +
\frac{g^2(r)}{C_0^2}  \, d\Omega_{(2)}^2  \label{le1}
\ee with
$C_0\neq 0$ and where
\begin{eqnarray}
g(r) &=& C_0\left( C_2+C_3r\right) -r^2 \,,\\
&&\nonumber\\
\rho(r) & = &  -5P(r) = \frac{5 \kappa \, C_0^2}{g^2(r)} \,.
\end{eqnarray}
In order to preserve the metric signature it must be $g(r)>0$ (if
$g(r)$ becomes negative, the coordinates $t$ and $r$ switch their
timelike and spacelike natures, as in the Schwarzschild geometry at
the horizon $r=2m$).

We rewrite the line element~(\ref{le1}) in terms of the areal radius
$R(r) = g(r)/|C_0|$. This relation is inverted by first obtaining
\be r^2 -C_0 C_3 r + \left( |C_0| R-C_0 C_2\right)=0 \ee and solving
for \be r (R) = \frac{1}{2} \left( C_0 C_3 \pm \sqrt{ C_0^2 C_3^2 +4
\left( C_0 C_2 -|C_0| R\right) } \right) \,. \ee The argument of the
square root in the right-hand side must be non-negative to keep $r$
real, which gives the limitation \be 0 \leq R <  \frac{ C_0^2
C_3^2+4 C_0 C_2}{4|C_0|} \equiv R_\mathrm{max} \label{Rmaxgu} \ee on
the range of the areal radius. The latter begins from zero at
$r_1=\frac{1}{2}\left(C_0 C_3-\sqrt{C_0^2 C_3^2+4C_0 C_2} \right)$,
increases to the maximum \be R_\mathrm{max} = R \left( \frac{C_0
C_3}{2} \right) \,, \ee and then decreases until it vanishes again
at $r_2=\frac{1}{2}\left(C_0 C_3+\sqrt{C_0^2 C_3^2+4C_0 C_2}
\right)$. The two coordinate charts $r_1\leq r \leq C_0 C_3/2$ and $
C_0 C_3/2 \leq r \leq r_2$ cover the same physical region $ 0\leq R
\leq R_\mathrm{max}$. We restrict ourselves to $r_1\leq r \leq C_0
C_3/2$, in which $dR/dr>0$, by choosing the negative sign in
Eq.~(2.5).

We write
\be
C_0 C_3-2r = \mp \sqrt{ C_0^2 C_3^2 +4\left( C_0
C_2 -|C_0|R\right)} \label{questa}
\ee
and, substituting the
relation between differentials
\be
dr= \frac{ |C_0|}{C_0C_3-2r} \, dR
\ee
and using Eq.~(\ref{questa}), the line element~(\ref{le1}) becomes
\be
ds^2 =  -\frac{|C_0|}{R} \, dt^2 + \frac{dR^2}{ 4\left(
\frac{R_\mathrm{max}}{R} -1\right)} +R^2 d\Omega_{(2)}^2 \,. \label{chosen}
\ee
The equation $\nabla^c R \nabla_c R =g^{RR}= 0 $ locating the
apparent horizons (see, {\em e.g.}, \cite{Faraoni:2015ula}) has
$R_\mathrm{max}$ as the only root, which is a single root and
therefore there are no apparent horizons for $R<R_\mathrm{max} $ (we
discuss the physical meaning of the formal root $ R_\mathrm{max} $
below).

The energy density \cite{Semiz:2020lxj} \be \rho(R) = \frac{5 \kappa
\, C_0^2}{g^2(r)}=  \frac{5\kappa}{R^2} \ee and the pressure
$P=-\rho/5$ (which are always non-zero) diverge at the origin $R =
0$, which corresponds to $r=r_1$, together with the Ricci scalar \be
{\cal R}= -\kappa \,  T = \kappa \left( \rho-3P \right)
=\frac{8\kappa}{5} \, \rho = \frac{8\kappa^2}{R^2} \,, \ee therefore
there is a naked spacetime singularity at the origin $R=0$.

The Misner-Sharp-Hernandez mass $M_\mathrm{MSH}(R)$  defined in
spherical symmetry by \cite{MSH1,MSH2}
\be
1-\frac{2M_\mathrm{MSH}}{R} = \nabla^c R \nabla_c R = g^{RR}
\ee
reads
\be
M_\mathrm{MSH} (R) = \frac{1}{2} \left( 5R-4R_\mathrm{max}\right)
\ee
for the geometry~(\ref{chosen}) and is negative in the region $ 0<R<
4R_\mathrm{max}/5 $ around the naked singularity. This fact is not
surprising: it has been argued that the Misner-Sharp-Hernandez mass (to
which the Hawking-Hayward quasilocal mass \cite{Hawking,Hayward} reduces
in spherical symmetry \cite{Haywardspherical}) is misbehaved near naked
singularities, Cauchy horizons, or regions with the wrong asymptotic
flatness \cite{Faraoni:2020stz,Faraoni:2020mdf}. This is the case, for
example, for the inner region of the Reissner-Nordstr\"om black hole near
the Cauchy horizon, for the entire Schwarschild spacetime with negative
mass, and for the Fisher-Janis–Newman–Winicour–Buchdahl–Wyman scalar field
solution of the Einstein equations \cite{Fisher:1948yn, BergmannLeipnik57,
Janis:1968zz, Buchdahl:1972sj, Wyman:1981bd, Dionysiu82,
Agnese:1985xj, Virbhadra:1997ie} for the parameter values for which
there is a naked singularity \cite{Faraoni:2021nhi}.

Let us come to the maximum value $R_\mathrm{max}$ of the areal radius
which, in spite of being a formal root of the equation $\nabla^cR \nabla_c
R=0$, does not describe a horizon but is instead the antipode of the origin
$R=0$ in a compact space.  To see this fact, it is instructive to study the
behaviour of radial null geodesics in this geometry. Consider the outgoing
$(+)$ and ingoing $(-)$ congruences of radial null geodesics with tangents
$l_{(\pm)}^{\mu} =dx^{\mu}/d\lambda$, where $\lambda$ is an affine parameter
along these curves. These tangents have components $l_{(\pm)}^\mu = \left( l^0,
l^1, 0, 0\right)$ and the normalization $l^{(\pm)}_{a} l_{(\pm)}^a=0$ yields
\be
l_{(\pm)}^1 = \pm\frac{2}{R}\sqrt{\left(
R_\mathrm{max}-R\right)|C_0|} \, l_{(\pm)}^0 \,;
\ee
since a null vector can be rescaled by a function, we can choose
$l^0 =1$ (which means choosing the coordinate time $t$ as the
affine parameter along these null geodesics),  obtaining
\begin{eqnarray}
l_{(\pm)}^\mu = \left(1, \pm\frac{2}{R}\sqrt{\left(
R_\mathrm{max}-R\right)|C_0|}, 0, 0 \right) \,.
\end{eqnarray}
We then have the first order equations
\begin{eqnarray}
&& \frac{dt}{d\lambda}=1\,,\\
&&\nonumber\\
&& \frac{R}{ \sqrt{R_\mathrm{max}-R}} \,\frac{d
R}{d\lambda}=\pm 2\sqrt{|C_0|}\,,
\end{eqnarray}
which integrate to
\begin{eqnarray}
&& t(\lambda)  = \lambda-\lambda_{0}\,,\\
&&\nonumber\\
&& \sqrt{R_\mathrm{max}-R}
\left( R +2R_\mathrm{max}\right) = \mp 3\sqrt{|C_0|}\left(
\lambda-\lambda_{0}\right)\,, \nonumber\\
&&
\end{eqnarray}
where $\lambda_0$ is an integration constant. Unfortunately this relation
cannot be inverted explicitly.

\begin{figure}
\includegraphics[scale=0.35]{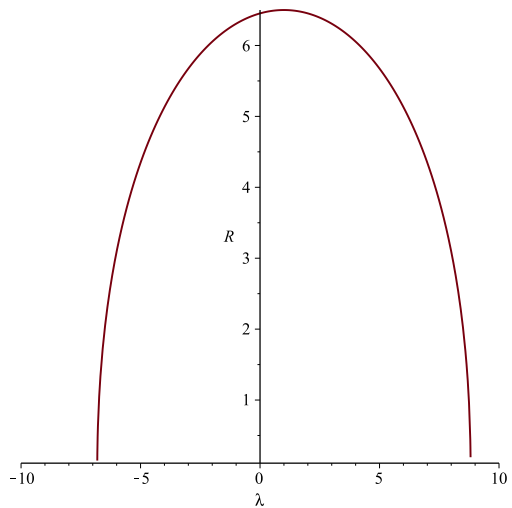}
\caption{\label{Fig:1} The areal radius $R $ versus the
affine parameter $\lambda$ along the radial null geodesics of the
geometry~(\ref{chosen}), for the parameter values $C_{0}= C_{2}=2$, $C_{3}=3$,
and $\lambda_{0}=1$. Outgoing
geodesics slow down as they approach $R_\mathrm{max}$, where they stop.
Ingoing geodesics starting near $R_\mathrm{max}$ do so extremely slowly
but accelerate as they approach $R=0$.  }
\end{figure}

Since
\be
\frac{dR}{dt} =\frac{dR}{d\lambda} = \pm 2\sqrt{|C_0|} \, \frac{ \sqrt{
R_\mathrm{max}-R}}{R}
\ee
(with the upper sign for outgoing and the lower one for ingoing radial
geodesics), near the origin $R=0$ it is $dR/dt \sim +\infty$ for outgoing
and $dR/dt \sim -\infty $ for ingoing geodesics. Furthermore, $dR/dt=0$ at
$R=R_\mathrm{max}$. Outgoing radial null geodesics starting near the
origin do so extremely fast but they slow down as they approach the
maximum possible radius $R_\mathrm{max}$, which can only be reached with
zero velocity (see Fig.~\ref{Fig:1}). A null geodesic starting exactly at
$R_\mathrm{max}$ does so with zero velocity $dR/d\lambda$ and remains
there. Ingoing radial null geodesics starting near the maximum radius
$R_\mathrm{max}$ are slow and accelerate as they get closer to the
central naked singularity, which they approach with infinite velocity
$dR/d\lambda\rightarrow -\infty$.

We can also study radial timelike geodesics with tangents $ u^{\mu}
=\left( u^0, u^1, 0, 0 \right)$. The normalization $u^c u_c=-1$ gives
\be
u^1 = \pm \frac{2}{\sqrt{R}} \sqrt{ \left( R_\mathrm{max}-R \right) \left[
\frac{ |C_0| (u^0 )^2 }{R} -1\right]} \,, \label{u1}
\ee
with the upper sign for outgoing and the lower one for ingoing geodesics.
The timelike Killing vector $\xi^a =\left( \partial/\partial t
\right)^a$ guarantees the conservation of the energy per unit mass of
the test particle $E$ along these geodesic curves:
\be
E =- g_{ab} \, \xi^a u^b = \frac{|C_0| u^0}{R} =\mbox{const.} \,,
\label{Killing}
\ee
where $u^0 >0 $ because these curves are future-oriented, hence $E$ is
strictly positive. Equation~(\ref{u1}) then gives

\be u^1 = \pm \, \frac{2}{\sqrt{R}} \, \sqrt{ \left(
R_\mathrm{max}-R\right) \left( \frac{E^2 R}{|C_0|} -1 \right)} \,,
\label{u1again} \ee
which tells us that:

\begin{itemize}

\item For a given energy $E$ determined by the initial condition $\left(
R_0, \dot{R}_0 \right)$, radial motion is only possible if \be R >
R_\mathrm{min}\equiv \frac{|C_0|}{E^2}  \ee (otherwise $u^1$ becomes
imaginary). Ingoing radial motion stops at $R_\mathrm{min}$ and a
test particle cannot approach the origin, which is consistent with
the fact that, according to Eq.~(\ref{Killing}), $u^0=ER/|C_0|
\rightarrow 0$ as $R\rightarrow 0$.

\item Outgoing radial motion stops at $R_\mathrm{max}$, where $u^1$
vanishes for
both outgoing and ingoing radial geodesics, and  a particle  starting
initially at $R_\mathrm{max}$ remains there irrespective of its
initial energy.

\item Since $R$ is limited by $ R_\mathrm{max}$, the possible energies are
bounded from below,

\be
E > \sqrt{ \frac{|C_0|}{ R_\mathrm{max} }} =
\frac{2 |C_0|}{ \sqrt{C_0^2 C_3^2 +4 C_0 C_2}}
 \equiv E_\mathrm{min} \,;
\ee particles with energy below, or equal to, the minimum threshold
$E_\mathrm{min}$ do not move.

\end{itemize}

\subsection{Case $C_2 \neq 0$, $C_3=0$}

In this case we are left with only two parameters $\left(
C_0, C_2 \right)$. Now  $ g(r) = C_0 C_2 -r^2 $, which requires
\be
C_0 C_2>0 \,, \quad\quad 0\leq r \leq \sqrt{C_0 C_2} \,.
\ee
The areal radius is
\be
R(r) =\frac{g(r)}{|C_0|}= \frac{C_0 C_2-r^2}{|C_0|}
\ee
with $r=0$ corresponding to $R=C_0C_2/|C_0|$,
while $ r=\sqrt{C_2 \, \mbox{sign}(C_0) } $ corresponds to the
origin $R=0$ of the physical radial coordinate.  The areal radius
$R(r)$  varies in the range
\be
0\leq R \leq C_2 \,\mbox{sign}(C_0) =|C_2|
\ee
(where, in the last equality, we used the fact that $C_0C_2>0$) and is a
decreasing function of $r$
since $dR/dr=-2r/|C_0| $ is always negative in the allowed range.
Inverting the relation between radial coordinates, one obtains
\be
r(R)= \sqrt{ C_0 C_2 -|C_0|R}
\ee
which, in conjunction with
\be
dr = -\frac{|C_0| dR}{2\sqrt{ C_0 C_2 -|C_0|R} }
\ee
yields the line element
\be
ds^2 = -\frac{|C_0|}{R} \, dt^2 +\frac{ dR^2}{4\left(
 \frac{|C_2|}{R}  - 1\right) } \, + R^2 d\Omega_{(2)}^2 \,.
\ee
This geometry is the same as that of the previous case $C_1=0$,
$C_3 \neq 0$ given by the line element~(\ref{chosen}), but now
$R_\mathrm{max}=|C_2|$. Again, the energy density is non-zero and the
Ricci scalar diverges at the origin $R=0$.

\subsection{Case $C_2=0 \,, C_3 \neq 0$}

For these parameter values, $g(r)= r\left( C_0 C_3 -r \right)$
requires $C_0 C_3$ to be positive and, therefore, we have the range
$ 0\leq r \leq C_0 C_3$ of the Buchdahl radius. Correspondingly, the
areal radius \be R(r) = \frac{r\left( C_0 C_3-r\right)}{|C_0|}
\label{questa2} \ee varies in the interval \be 0 \leq R \leq \frac{
|C_0| C_3^2}{4}  \,, \ee beginning from zero at  $r=0$, increasing
to the maximum \be R_\mathrm{max}  \equiv R \left( \frac{C_0 C_3}{2}
\right)  = \frac{ |C_0| C_3^2}{4} \,, \label{Rmaxsu} \ee and then
decreasing until it vanishes again at $r=C_0 C_3$. There are two
coordinate charts $0\leq r \leq C_0 C_3/2$ and $ C_0 C_3/2 \leq r
\leq C_0 C_3$ covering the same physical region $0 \leq R \leq
R_\mathrm{max}$ and we restrict ourselves to the former, in which
$dR/dr>0$. Equation~(\ref{questa2}) yields
\be
r^2 -C_0C_3 r
+|C_0|R=0 \ee with roots \be r(R) =\frac{1}{2} \left( C_0 C_3 \pm
\sqrt{ C_0^2 C_3^2 -4|C_0|R} \, \right) \,,
\ee
where we choose the lower sign for consistency with $dR/dr>0$ and $ 0\leq
r\leq C_0C_3/2$. Then $ g(r)=|C_0|R $ and
\be
dr=\frac{|C_0|}{\sqrt{ C_0^2 C_3^2 -4|C_0|R}} \, dR
\ee
give the line element
\be
ds^2 = -\frac{|C_0|}{R} \, dt^2 + \frac{dR^2}{4\left(  \frac{
R_\mathrm{max} }{R} - 1 \right) } \, +R^2 d\Omega_{(2)}^2
\ee
which is the same as the line element~(\ref{chosen}), but with
$R_\mathrm{max}$ now given by Eq.~(\ref{Rmaxsu}). The Ricci scalar
\be {\cal R} =\frac{8\kappa}{5} \, \rho = \frac{8\kappa^2}{R^2} =
\frac{8\kappa^2 C_0^2}{r^2\left( C_0 C_3-r\right)^2 } \,, \ee
diverges at the origin $R=0$ (which corresponds to  $r=0$ in the
chart with $dR/dr>0$), therefore there is a naked spacetime
singularity there.

\section{General solution $C_1\neq 0$}
\label{sec:3} \setcounter{equation}{0}

The line element for the generic family of Semiz solutions
is~(\ref{generalle}) \cite{Semiz:2020lxj}. The presence of four
parameters with relatively wide ranges now makes it difficult to reach
definite conclusions and we focus on special cases.

\subsection{$C_3=0$ is Schwarzschild}

When $C_3 =0$, the energy density~(\ref{rho}) and the pressure
$P=-\rho/5 $ vanish identically and this spacetime is empty. Since
the geometry is also spherically symmetric and asymptotically flat
(as we are going to show) it must be the Schwarzschild one,
according to the Jebsen-Birkhoff theorem \cite{Waldbook}. In fact,
we have $ f(r) = 3\left(C_1 C_2+r \right) $, the areal radius is
\be
R=\frac{C_1C_2 +r}{|C_1|} \,, \ee and \be C_0+C_1r= C_1 |C_1|R
+C_0-C_1^2 C_2 \,,
\ee
then $dr=|C_1|dR$, yielding the line element
\begin{eqnarray}
ds^2 &=&-\frac{ C_1^2 R +\left( C_0 -C_1^2 C_2\right)
\,\mbox{sign}(C_1)}{R}
\, dt^2 \nonumber\\
&&\nonumber\\
&\, &  + \frac{C_1^2R}{C_1^2 R +\left( C_0-C_1^2 C_2\right)
\,\mbox{sign}(C_1)}
\, dR^2 +R^2 d\Omega_{(2)}^2 \nonumber\\
&&\label{lequesto}\\
&\simeq &  -d\bar{t}^2 +dR^2 +R^2 d\Omega_{(2)}^2 \quad
\quad \mbox{as} \: R\rightarrow + \infty \,,
\end{eqnarray}
where $d\bar{t} \equiv |C_1| dt$. This geometry is
asymptotically flat: by introducing the constant
\be
m \equiv
\frac{1}{2C_1^2}\left( C_1^2 C_2-C_0\right) \, \mbox{sign}(C_1)
\ee
(which is not necessarily positive) and rescaling the time
coordinate according to $t \rightarrow \bar{t} = |C_1| \, t$, the
line element~(\ref{lequesto}) is written as the Schwarzschild one
\be
ds^2=-\left(1-\frac{2m}{R} \right) d\bar{t}^2
+\frac{dR^2}{1-2m/R} +R^2 d\Omega_{(2)}^2
\ee
describing a black hole if $m>0$ and a naked central singularity if $m<0$.

\subsection{Special case $C_2=0$}

We have three parameters $\left( C_0, C_1, C_3 \right)$ with
$C_1C_3 \leq 0$ and  now $f(r)= 3r+C_3\left( C_0 + C_1 r\right)^3$;
the areal radius is
\be
R(r)= \frac{ 3r +C_3\left( C_0+C_1 r
\right)^3}{3|C_1|} \,.
\ee
We have
\be
\frac{dR}{dr}= \frac{1}{|C_1|} \left[ 1- |C_1C_3| \left( C_0 +C_1r \right)^2
\right] \,,
\ee
which is positive for
\be
\left| r+\frac{C_0}{C_1} \right| < \frac{1}{|C_1| \sqrt{ |C_1 C_3|}} \,.
\ee
To proceed, let us consider the situation $r\geq -C_0/C_1$, in which case $R$
increases in the interval
\be
r_\mathrm{min} \equiv -\frac{C_0}{C_1} \leq r \leq \frac{1}{|C_1| \sqrt{
|C_1C_3|}} -\frac{C_0}{C_1} \equiv r_\mathrm{max}
\ee
with $ R_\mathrm{min} \leq R \leq R_\mathrm{max} $ and
\begin{widetext}
 \begin{eqnarray}
R_\mathrm{min} & \equiv & R\left(r_\mathrm{min} \right) =
\frac{-C_0}{C_1|C_1|}\,,\\
&&\nonumber\\
R_\mathrm{max}  & \equiv & R \left( r_\mathrm{max} \right) =
\frac{1}{3|C_1|}
\left[ \frac{3}{|C_1| \sqrt{|C_1 C_3|}}-\frac{ 3C_0}{C_1} +  C_3
\left( \frac{ \mbox{sign}(C_1)}{ \sqrt{|C_1 C_3|} } \right)^3
\right] \nonumber\\
&&\nonumber\\
&=&  \frac{1}{3|C_1|}
\left[ \frac{3 + \, \mbox{sign}(C_1 C_3)  }{|C_1| \sqrt{|C_1 C_3|}
}-3\frac{
C_0}{C_1} \right]     \nonumber\\
&&\nonumber\\
&=& \left\{ \begin{array}{lll}
\frac{1}{3|C_1|} \left( \frac{2}{|C_1| \sqrt{|C_1 C_3|}} -\frac{3C_0}{C_1}
\right) & \quad \mbox{if} & C_1 C_3<0 \,,\\
&&\\
\frac{1}{C_1^2} \left[ \frac{1}{\sqrt{|C_1 C_3|}} -C_0 \, \mbox{sign}(C_1)
\right] & \quad \mbox{if} & C_1 C_3 =0 \,.
\end{array} \right.
\end{eqnarray}
\end{widetext}
We have again a compact space. Rewriting the line element
(\ref{generalle}) in terms of the areal radius produces a cumbersome
expression that does not depend only on $R$ but contains also $r(R)$
because the relation $R(r)$ cannot be inverted explicitly.

\subsection{The even more special case $C_0=C_2=0$}

In this case we have only two parameters $\left( C_1, C_3 \right)$, $ f(r)=
r\left( 3+C_1^3 C_3 r^2 \right)$, and the areal radius is
\begin{eqnarray}
R(r) &=& \frac{f(r)}{3|C_1|}=\frac{ 3r+ C_1^3 C_3 r^3}{3|C_1|} \nonumber\\
&&\nonumber\\
& = &  \frac{r \left[ 3+ (C_1 C_3)C_1^2r^2 \right] }{3|C_1| }
\leq \frac{r}{|C_1|}    \,, \label{eq:stukaz}
\end{eqnarray}
where the last inequality follows from $C_1 C_3 \leq 0$. Since
\be
\frac{dR}{dr} =\frac{ 1- |C_1 C_3| C_1^2 r^2}{|C_1|} \geq
0  \quad \forall r\in \left(0, \frac{1}{ |C_1| \sqrt{ |C_1 C_3|} }
\right) \,,
\ee
the areal radius  is an increasing function of $r$
in the interval $\left(0, \frac{1}{|C_1| \sqrt{ |C_1 C_3|}} \right)
$ with $R(0)=0$, is maximum at $ \frac{1}{ |C_1| \sqrt{ |C_1 C_3|}}
 $ and then decreases, vanishing
again at $r= \sqrt{ \frac{3}{|C_1 C_3| C_1^2}}$. This compact space
corresponds to the range
\be
0  \leq R \leq R_\mathrm{max} = \frac{2}{3C_1^2  \sqrt{ |C_1C_3|} }
\ee
of the  areal radius, with $
R \simeq r/|C_1| $  as $ r\rightarrow 0^{+}$.
Equation~(\ref{eq:stukaz}) is inverted by first obtaining \be C_3
C_1^3 r^3+3r-3|C_1|R=0 \ee and then solving for
\be
r= \frac{ \left[
A(R)\right]^{1/3} }{2 C_3 C_1^2}-\frac{2}{C_1 \left[
A(R)\right]^{1/3} } \ee where \be A(R)=\left(12 |C_1|C_1 R + 4\sqrt{
9 C_1^4  R^2 +4 } \, \right) C_1^2 C_3^2 \,,
\ee
while the two remaining roots are imaginary. Substituting the relation
between differentials
\be
dr=\frac{|C_1|}{1+C_3 C_1^3 r^2} \, dR
\ee
and
using
\be
1 + C_1^3  C_3 r^2=\frac{ \left[ A(R)\right]^{2/3}}{4C_1
C_3}+\frac{4C_1 C_3}{\left[ A(R)\right]^{2/3}}-1 \ee yield the line
element
\begin{eqnarray}
ds^2 & =& -\frac{1}{|C_1| R}\left\{ \frac{ \left[ A(R)\right]^{1/3}
}{2C_3} - \frac{2C_1}{\left[ A(R)\right]^{1/3} } \right\} dt^2
\nonumber\\
&&\nonumber\\
&\, & + \frac{C_1^2|C_1|R}{B(R)} \, dR^2  + R^2 d\Omega^2_{(2)} \,,
\end{eqnarray}
where
\begin{eqnarray}
B(R) &=& \frac{ \left[ A(R) \right]^{5/3}}{32 C_1^2 C_3^3 }-\frac{32
C_1^3 C_3^2 }{ \left[ A(R)\right]^{5/3} } + \frac{5 \left[
A(R)\right]^{1/3}
}{2C_3}-\frac{3A(R) }{8 C_1 C_3^2} \nonumber\\
&&\nonumber\\
&\, & + \frac{24 C_1^2 C_3 }{A(R)}  -\frac{10 C_1}{ \left[ A(R)
\right]^{2/3}} \,.
\end{eqnarray}
Again, the many combinations of parameters and the cumbersome
metric coefficients do not lend themselves to a straightforward and
transparent analysis, but it is clear that also in this case we have a
compact 3-space of finite extent.

Using (3.20), the energy density~(\ref{rho}) reduces to
\begin{eqnarray}
\rho(R) &=& -5P(R)= \frac{5 {\cal R}}{8\kappa} =   - \frac{5\kappa \,
C_1^3 C_3 r^2}{R^2}  \nonumber\\
&&\nonumber\\
& \approx &  - \frac{5\kappa}{R^2} \left[(C_1  C_3)^{1/3}
+\frac{1}{(C_1  C_3)^{1/3} }  -2  \right] \,,\nonumber\\
&&
\end{eqnarray}
as $R\rightarrow 0^{+}$. Hence, $\rho$ and $P$ are singular at the
origin, together with the Ricci scalar ${\cal R}$ and \be {\cal
R}_{ab} {\cal R}^{ab} = \frac{28 \kappa^2}{25} \, \rho^2 \,. \ee As
$R\rightarrow 0^{+}$, we have the asymptotics
\begin{eqnarray}
A(R) & \approx & 8C_1^2 C_3^2 \,,\\
&&\nonumber\\
B(R) & \approx & \left( C_1^4 C_3 \right)^{1/3} -\frac{7}{2 \left(
C_1 C_3^4 \right)^{1/3} } + 5 \left(\frac{C_1^2}{C_3} \right)^{1/3} -3C_1
\nonumber\\
&&\nonumber\\
&\, & +\frac{3}{C_3}  \equiv B_0 \,,\\
&&\nonumber\\
g_{00} &\approx & -\frac{1}{|C_1|R} \left[
\left( \frac{C_1^2}{C_3} \right)^{1/3} -
\left( \frac{C_1}{C_3^2} \right)^{1/3} \right] \,,\\
&&\nonumber\\
 g_{11} & \approx & \frac{ C_1^2 |C_1| R}{B_0} \,,
\end{eqnarray}
and $g_{00} \rightarrow \infty$ while $g_{11} \rightarrow 0$ as
$R\to 0$.

\section{Conclusions}
\label{sec:5}
\setcounter{equation}{0}

We have studied the nature of the new classes of static and spherically
symmetric solutions of the Einstein equations given recently in
Ref.~\cite{Semiz:2020lxj} when the matter source is a perfect fluid with
equation of state $P=-\rho/5$. The analytical solutions of
Ref.~\cite{Semiz:2020lxj} that we analyzed (except for the Schwarzschild
solution obtained for $C_3=0$) describe compact spaces with naked central
singularities. The ``general'' family of solutions~(\ref{generalle})
and~(\ref{rho}) always reduces to Schwarschild when the parameter $C_3$
vanishes. In most other situations, the presence of three or four
parameters and/or the cubic nature of the function $R(r)$ hamper a
complete description of the geometry. However, in all cases analyzed,
except for the empty spacetime associated with $C_3=0$, we find a compact
space of finite volume (a feature mentioned in \cite{Semiz:2020lxj}).

The fact that the geometry, together with the energy density and the
pressure, is singular at $R=0$ is not necessarily the death knell for
these solutions. In fact, it is deemed acceptable for fluid solutions of
the Einstein equations to only model limited regions of relativistic
stars, a procedure that is reflected in the authoritative
Ref.~\cite{Stephani} and in the more specialized literature. Indeed, even
Newtonian stars are rarely modelled with a single fluid, corresponding to
the fact that different regions at different temperatures and densities
are described by different equations of state unless the stellar material
is well mixed, which only happens in certain types of stars. Therefore,
there is in principle the (physically well motivated) possibility of
excising the singularity and replacing it with a more realistic geometry
sourced by matter with a different equation of state. However, if
one wants to describe a stellar interior with this exotic fluid, one must
match it with an asymptotically flat Schwarzschild exterior. The fact that
the solutions of \cite{Semiz:2020lxj} describe compact spaces points to a
possible analogy with the Oppenheimer-Snyder model of gravitational
collapse to a black hole \cite{OppenheimerSnyder}. In this model, a
compact, positively curved
Friedmann-Lema\^itre-Robertson-Walker universe collapsing to a Big Crunch
is matched to a Schwarzschild exterior on the surface of a 2-sphere of
symmetry \cite{OppenheimerSnyder}, satisfying the Darmois-Israel junction
conditions \cite{Darmois,Israel}. However, in the Oppenheimer-Snyder model
the matching is possible because the collapsing interior universe is
filled by a dust with zero pressure everywhere. It is well known that the
matching to a Schwarzschild exterior can only be done on a surface on
which the pressure $P(R)$ vanishes, otherwise the junction conditions are
violated and there is a material layer on the matching surface, which is
certainly not an ingredient of realistic stellar models. (This fact is
highlighted in many studies of relativistic fluid balls
\cite{Vaidyaball, MashhoonPartovi, SrivastavaPrasad, ThompsonWhitrow,
ThompsonWhitrow2, Bondi, BondiNature, Faraoni:2020uuf}  and fireballs
\cite{SmollerTemple}.)
However, for the fluid solutions of \cite{Semiz:2020lxj} under discussion,
the pressure $P(R)$ never vanishes. Therefore, the best that one could do
is modelling a limited region of a stellar interior with the Semiz
solutions for $P=-\rho/5$. To be physical, this region should correspond
to a positive Misner-Sharp-Hernandez mass $M_\mathrm{MSH}$ and, therefore,
should be sufficiently far away from the singularity at $R=0$. The excised
region containing the origin should be modelled with a different,
non-singular, solution of the Einstein equations.\footnote{There is a
subtlety: matching a Semiz region with another fluid solution is in
principle possible because the Semiz geometries solve the Einstein
equations with zero cosmological constant $\Lambda$. A non-zero $\Lambda
$ would be present in all spacetime regions and matching one of them with
a Schwarzschild exterior is impossible \cite{Faraoni:2021vpn}.} Then, the
$w=-1/5$ solution should be matched continuously with another
``intermediate'' solution with non-vanishing pressure on a surface of
constant radius, and the pressure in this layer should then go to zero at
larger radii to make it possible to match it to a Schwarzschild exterior,
satisfying again the Darmois-Israel junction conditions. In the absence of
a specific need for such an involved ``star'' model in astrophysics, we
will not pursue this object further, limiting ourselves to pointing out
the constraints for such a construction. Probably some of the
phenomenology unveiled here for the geometries found in
\cite{Semiz:2020lxj} also applies to other classes of perfect fluid
solutions of the Einstein equations. Whether this is the case will be
established in future work.

\begin{acknowledgments}

This work is supported, in part, by the Natural Sciences \&
Engineering Research Council of Canada (grant no. 2016-03803 to
V.F.).

\end{acknowledgments}



\begin{thebibliography}{99}

\bibitem{Semiz:2020lxj} \.I.~Semiz, ``The general static spherical perfect
fluid solution with EoS parameter $w=-1/5$,'' [arXiv:2007.08166
[gr-qc]].

\bibitem{DeBenedictis:2008qm} A.~DeBenedictis, R.~Garattini and
F.~S.~N.~Lobo, ``Phantom stars and topology change,'' Phys. Rev. D
\textbf{78}, 104003 (2008) doi:10.1103/PhysRevD.78.104003 [arXiv:0808.0839
[gr-qc]].

\bibitem{Chapline:2004jfp} G.~Chapline, ``Dark energy stars,'' eConf
\textbf{C041213}, 0205 (2004) [arXiv:astro-ph/0503200 [astro-ph]].

\bibitem{Lobo:2005uf} F.~S.~N.~Lobo, ``Stable dark energy stars,'' Class.
Quant. Grav. \textbf{23}, 1525-1541 (2006)
doi:10.1088/0264-9381/23/5/006 [arXiv:gr-qc/0508115 [gr-qc]].

\bibitem{Bilic:2005sn} N.~Bilic, G.~B.~Tupper and R.~D.~Viollier,
``Born-infeld phantom gravastars,'' JCAP \textbf{02}, 013 (2006)
doi:10.1088/1475-7516/2006/02/013 [arXiv:astro-ph/0503427
[astro-ph]].

\bibitem{Chan:2008rk} R.~Chan, M.~F.~A.~da Silva and J.~F.~Villas da
Rocha, ``Star Models with Dark Energy,'' Gen. Relativ. Gravit.
\textbf{41}, 1835-1851 (2009) doi:10.1007/s10714-008-0755-9
[arXiv:0803.3064 [gr-qc]].

\bibitem{Yazadjiev:2011sm} S.~S.~Yazadjiev, ``Exact dark energy star
solutions,'' Phys. Rev. D \textbf{83}, 127501 (2011)
doi:10.1103/PhysRevD.83.127501 [arXiv:1104.1865 [gr-qc]].

\bibitem{Rahaman:2011hd} F.~Rahaman, R.~Maulick, A.~K.~Yadav, S.~Ray and
R.~Sharma, ``Singularity-free dark energy star,'' Gen. Relativ. Gravit.
\textbf{44}, 107-124 (2012) doi:10.1007/s10714-011-1262-y [arXiv:1102.1382
[gr-qc]].

\bibitem{Horvat:2012aq} D.~Horvat and A.~Marunovi\'c, ``Dark energy-like
stars from nonminimally coupled scalar field,'' Class. Quant. Grav.
\textbf{30}, 145006 (2013) doi:10.1088/0264-9381/30/14/145006
[arXiv:1212.3781 [gr-qc]].

\bibitem{Bhar21} P. Bhar, ``Dark energy stars in Tolman–Kuchowicz
spacetime in the context of Einstein gravity'', Phys. Dark Univ.
{\bf 34}, 100879 (2021).

\bibitem{Armendariz-Picon:2005oog} C.~Armendariz-Picon and E.~A.~Lim,
``Haloes of k-essence,'' JCAP \textbf{08}, 007 (2005)
doi:10.1088/1475-7516/2005/08/007 [arXiv:astro-ph/0505207 [astro-ph]].

\bibitem{AmendolaTsujikawabook} L. Amendola and S. Tsujikawa, {\it Dark
Energy: Theory and Observations} (Cambridge University Press, Cambridge,
UK, 2010).

\bibitem{Stephani} H. Stephani, D. Kramer, M. MacCallum, C. Hoenselaers,
E. Herlt, {\em Exact Solutions of the Einstein Field Equations}
(Cambridge University Press, Cambridge, 2003).

\bibitem{Delgaty:1998uy} M.~S.~R.~Delgaty and K.~Lake, ``Physical
acceptability of isolated, static, spherically symmetric, perfect fluid
solutions of Einstein's equations,'' Comput. Phys. Commun. \textbf{115},
395-415 (1998) doi:10.1016/S0010-4655(98)00130-1 [arXiv:gr-qc/9809013
[gr-qc]].

\bibitem{Waldbook} R.~M. Wald, {\em General Relativity} (Chicago
University Press, Chicago, 1984).

\bibitem{Faraoni:2015ula} V.~Faraoni, {\it Cosmological and Black Hole
Apparent Horizons}, Lect. Notes Phys. \textbf{907} (Springer, New York,
2015) doi:10.1007/978-3-319-19240-6

\bibitem{MSH1} C.~W. Misner and D.~H. Sharp, ``Relativistic
Equations for Adiabatic, Spherically Symmetric Gravitational Collapse'',
Phys. Rev. {\bf 136}, B571 (1964).

\bibitem{MSH2} W.~C. Hernandez and C.~W. Misner, ``Observer time as a
coordinate in relativistic spherical hydrodynamics'', Astrophys. J. {\bf
143}, 452 (1966).

\bibitem{Hawking} S.~W. Hawking, `` Gravitational radiation in an
expanding
universe'', J. Math. Phys. {\bf 9}, 598 (1968).

\bibitem{Hayward} S.~A. Hayward, ``Quasilocal gravitational energy'',
Phys. Rev. D {\bf 49}, 831 (1994).

\bibitem{Haywardspherical} S.~A. Hayward, ``Gravitational energy in
spherical symmetry'', Phys. Rev. D {\bf 53}, 1938 (1996).

\bibitem{Faraoni:2020stz} V.~Faraoni and A.~Giusti, ``Unsettling physics
in the quantum-corrected Schwarzschild black hole,'' Symmetry \textbf{12},
no.8, 1264 (2020) doi:10.3390/sym12081264 [arXiv:2006.12577 [gr-qc]].

\bibitem{Faraoni:2020mdf} V.~Faraoni, A.~Giusti and T.~F.~Bean,
``Asymptotic flatness and Hawking quasilocal mass,'' Phys. Rev. D
\textbf{103}, no.4, 044026 (2021) doi:10.1103/PhysRevD.103.044026
[arXiv:2010.00069 [gr-qc]].

\bibitem{Fisher:1948yn} I.~Z.~Fisher, ``Scalar mesostatic field with
regard for gravitational effects,'' Zh. Eksp. Teor. Fiz. \textbf{18},
636-640 (1948) [arXiv:gr-qc/9911008 [gr-qc]].

\bibitem{BergmannLeipnik57} O. Bergmann and R. Leipnik, ``Space-time
structure of a static spherically symmetric scalar field'', Phys. Rev. 107
(1957) 1157–1161, http: //dx.doi.org/10.1103/PhysRev.107.1157.

\bibitem{Janis:1968zz} A.~I.~Janis, E.~T.~Newman and J.~Winicour,
``Reality of the Schwarzschild Singularity,'' Phys. Rev. Lett.
\textbf{20}, 878-880 (1968) doi:10.1103/PhysRevLett.20.878

\bibitem{Buchdahl:1972sj} H.~A.~Buchdahl, ``Static solutions of the
Brans-Dicke equations,'' Int. J. Theor. Phys. \textbf{6}, 407-412 (1972)
doi:10.1007/BF01258735

\bibitem{Wyman:1981bd} M.~Wyman, ``Static Spherically Symmetric Scalar
Fields in General Relativity,'' Phys. Rev. D \textbf{24}, 839-841 (1981)
doi:10.1103/PhysRevD.24.839

\bibitem{Dionysiu82} D. D. Dionysiou, ``Static spherically-symmetric
scalar-field theory in general relativity'', Astrophys. Space Sci. {\bf
88}, 493 (1982).

\bibitem{Agnese:1985xj} A.~G.~Agnese and M.~La Camera, ``Gravitation
without black holes'', Phys. Rev. D \textbf{31}, 1280-1286 (1985)
doi:10.1103/PhysRevD.31.1280

\bibitem{Virbhadra:1997ie} K.~S.~Virbhadra, ``Janis-Newman-Winicour and
Wyman solutions are the same,'' Int. J. Mod. Phys. A \textbf{12},
4831-4836 (1997) doi:10.1142/S0217751X97002577 [arXiv:gr-qc/9701021
[gr-qc]].

\bibitem{Faraoni:2021nhi} V.~Faraoni, A.~Giusti and B.~H.~Fahim,
``Spherical inhomogeneous solutions of Einstein and
scalar\textendash{}tensor gravity: A map of the land,'' Phys. Rept.
\textbf{925}, 1-58 (2021) doi:10.1016/j.physrep.2021.04.003
[arXiv:2101.00266 [gr-qc]].

\bibitem{OppenheimerSnyder} J.~R. Oppenheimer and J.~R. Snyder, ``On
continued gravitational contraction'', Phys. Rev. {\bf 56}, 455 (1939).

\bibitem{Darmois} G. Darmois, ``Les Equations de la Gravitation
Einsteinienne'', {\em Memorial des Sciences Mathematiques}
XXV (Gauthier-Villars, Paris, 1927).

\bibitem{Israel} W. Israel, ``Singular hypersurfaces and thin shells in
general relativity'', Nuovo Cimento~B {\bf 44}, 1 (1966); {\em Errata}
{\bf 48}, 463(E) (1967).

\bibitem{Vaidyaball} P.~C. Vaidya, ``Nonstatic Analogs of Schwarzschild's
Interior Solution in General Relativity'', Phys. Rev. {\bf 174}, 1615
(1968).

\bibitem{MashhoonPartovi} B. Mashhoon and M.~H. Partovi, ``On the
gravitational motion of a fluid obeying an equation of state'', Ann. Phys.
(NY) {\bf 130}, 99 (1980).

\bibitem{SrivastavaPrasad} D.~C. Srivastava and S.~S. Prasad, ``Perfect
Fluid Spheres in General Relativity'', {\em Gen. Relativ. Gravit.} {\bf
15}, 65 (1983).

\bibitem{ThompsonWhitrow} A.~H. Thompson and W.~J. Whitrow,
``Time-Dependent Internal Solutions for Spherically Symmetrical Bodies in
General Relativity: I. Adiabatic Collapse'', Mon. Not. R. Astron. Soc.
{\bf 136}, 207 (1967).

\bibitem{ThompsonWhitrow2} A.~H. Thompson and W.~J. Whitrow,
``Time-Dependent Internal Solutions for Spherically Symmetrical Bodies in
General Relativity: II. Adiabatic Radial Motions of Uniformly Dense
Spheres'', Mon. Not. R. Astron. Soc. {\bf 139}, 499 (1968).

\bibitem{Bondi} H. Bondi, ``Gravitational bounce in general relativity'',
Mon. Not. R. Astron. Soc. {\bf 142}, 333 (1969).

\bibitem{BondiNature} H. Bondi, ``Bouncing Spheres in General
Relativity'', {\em Nature} {\bf 215}, 838 (1967).

\bibitem{Faraoni:2020uuf} V.~Faraoni and F.~Atieh, ``Turning a Newtonian
analogy for FLRW cosmology into a relativistic problem,'' Phys. Rev. D
\textbf{102}, no.4, 044020 (2020) doi:10.1103/PhysRevD.102.044020
[arXiv:2006.07418 [gr-qc]].

\bibitem{SmollerTemple} J. Smoller and B. Temple, ``Shock-wave solutions
in closed form and the Oppenheimer-Snyder limit in General Relativity'',
SIAM J. Appl. Math. {\bf 58}, 15 (1998).

\bibitem{Faraoni:2021vpn} V.~Faraoni, S.~Jose and A.~Leblanc, ``Curious
case of the Buchdahl-Land-Sultana-Wyman-Iba\~nez-Sanz spacetime,'' Phys.
Rev. D \textbf{105}, no.2, 024030 (2022) doi:10.1103/PhysRevD.105.024030
[arXiv:2110.11289 [gr-qc]].

\end{thebibliography}

\end{document}